# Simulative Comparison of DVB-S2X/RCS2 and 3GPP 5G NR NTN Technologies in a Geostationary Satellite Scenario

Lauri Sormunen, Tuomas Huikko, Verneri Rönty, Erno Seppänen, Sami Rantanen, Frans Laakso, Vesa Hytönen, Mikko Majamaa and Jani Puttonen
Magister Solutions Ltd.
Jyväskylä, Finland
{firstname.lastname}@magister.fi

*Abstract*—Comparison between existing, well-established satellite technologies, like the Digital Video Broadcasting (DVB) satellite specifications, and the emerging Third Generation Partnership Project (3GPP) specified 5th Generation New Radio (5G NR) Non-Terrestrial Networks (NTN) is an actively discussed topic in the satellite industry standardization groups. This article presents a thorough performance comparison between DVB Second Generation Satellite Extensions (DVB-S2X) and Return Channel via Satellite 2nd Generation (DVB-RCS2), and NR NTN in a Geostationary Orbit (GEO) satellite scenario, using system-level simulators (SLS) for evaluation, namely Satellite Network Simulator 3 (SNS3) and ALIX 5G (TN-)NTN SLS, built on the same Network Simulator 3 (ns-3) platform. With the satellite system geometry, beam layout, and link budget aligned to use the 3GPP NTN example parameterization for a fair comparison between DVB and NR NTN, the results show that DVB-S2X consistently achieves higher spectral efficiency than the NR Physical Downlink Shared Channel (PDSCH) on the forward user link. In contrast, on the return link, the NR Physical Uplink Shared Channel (PUSCH) demonstrates better spectral efficiency at the system level. The SLS results incorporate link-level performance, obtained through link-level simulations (LLS) for different modulation and coding schemes (MCS) and waveforms supported by each technology.

I. INTRODUCTION

This article presents a simulative comparison of 5th Generation (5G) New Radio Non-Terrestrial Networks (NR NTN) and Digital Video Broadcasting (DVB) Second Generation Satellite Extensions (S2X) on the forward user link (Downlink, DL), and DVB Return Channel via Satellite 2nd generation (RCS2) on the return user link (Uplink, UL), with a specific focus on Geostationary Orbit (GEO) satellites. Operating within the 20-30 GHz frequencies, commonly known as the Ka-band, or Frequency Range 2 (FR2) in 3rd Generation Partnership Project (3GPP) terminology, this paper provides a comprehensive performance assessment of the air interface technologies in a GEO, Ka-band system.

The NR NTN technology is an extension of the more general 5G NR standard, which has its roots in terrestrial cellular networks. The standardization effort of NR is driven by 3GPP, and the initial extension to NTN was originally fully specified in 3GPP Release-17 [1], where a large portion of the current NTN contributions were reported in 3GPP Technical Reports 38.811 [2] and 38.821 [3].

The improvement effort of NR in NTN, and NR in general, is currently ongoing and requires supporting research to create updated specifications.

The DVB-S2(X) [4][5] and RCS2 [6][7][8] specifications, on the other hand, were initially released in 2014 and 2012, respectively, and are still actively being maintained and updated by the DVB project consortium. Since their release, the DVB-S2X and RCS2 specifications have been widely adopted across various market segments. Originally developed for satellite television broadcasting, the DVB satellite specifications have evolved to accommodate a broader range of use cases. This evolution has enhanced their relevance to modern needs, building on a solid foundation in satellite communication.

The primary aim of this article is to compare the performance of broadband satellite services operating in geostationary orbits using these two alternative technologies. In doing so, the intent is to provide valuable insights that can support standardization efforts within the European Telecommunications Standards Institute (ETSI) and 3GPP. This endeavor aligns closely with the ETSI SES-SCN working group's work item, "Comparison of DVB-S2X/RCS2 and 3GPP 5G-NR NTN based systems for broadband satellite communication systems" [9], for which the results of this article have also been contributed. The work item initially provided an analytical comparison; however, the simulation-based comparison aims to produce more practical and applicable results, for a more effective evaluation. It is important to also acknowledge previous research efforts comparing the two technologies, where the comparison has been typically either limited to link level/waveform performance analysis [10] or system level analysis in a highly limited scope [11]. This article aims for a more in-depth comparison, compared to previous works, for a more comprehensive system level evaluation, with the help of thorough link level evaluation performed in [12]. The analytical comparison within the report found that the difference in the originally intended use cases of the technologies resulted in some differences in the current standards. Regarding functional aspects, NR has better support for e.g., mobility, energy saving, network slicing, multi-connectivity and the backhaul service aspects. Operationally, both technologies can support similar satellite systems (both Low Earth Orbit (LEO) and GEO), operating in Ka- or Ku-band under similar operational constraints. In terms of performance, NR was found to have slightly higher access and physical layer overheads on the forward link due to framing and signaling on both layers, while overheads were slightly lower on the return link. Link level signal degradation from Peak-to-Average Power Ratio (PAPR) was estimated to remain similar, if Single Carrier Frequency Division Multiple Access (SC-FDMA) is utilized in the UL [13], and in DL, if multiple carriers are multiplexed within the same high-power amplifier [14].

## II. SIMULATOR DESCRIPTIONS

This section describes the simulation tools used to generate results for comparing NR NTN and DVB-S2X/RCS2 technologies.

### A. DVB-S2X/RCS2 System Level Simulator - SNS3

Satellite Network Simulator 3 (SNS3) is a satellite network extension to the Network Simulator 3 (ns-3) platform [15]. ns-3 is an open-source, discrete-event simulator designed for research and educational use, providing a common C++ framework for developing packet-level simulators across various technologies and licensed under the General Public License v2 (GPLv2). The physical/link layer performance of SNS3 is abstracted underneath a Link-to-System (L2S) mapper and a set of modulation and coding (MODCOD) schemes. Specific signal-to-interference-plus-noise Ratio (SINR) to Frame Error Rate (FER) mapping curves have been adapted from link-level simulation (LLS) results. ns-3 can be categorized as a Non-Real Time (NRT) network level system simulator, which models the protocols from physical layer up to application layer in quite high level of accuracy. SNS3 models a fully interactive multi-spot beam satellite network with a geostationary satellite and transparent (bent pipe) payload. SNS3 was originally developed in a European Space Agency (ESA) project [16][17], and an open-source version of the simulator is also available [18]. The air interface architecture of SNS3 is illustrated in Fig. 1.

SNS3 has been developed to comply with the DVB-S2(X) [4][5] and DVB-RCS2 [7] specifications. SNS3 also supports comparison with the NR NTN standard by implementing Bessel beam patterns, antenna models [2] and Frequency Reuse Factor (FRF) schemes [3] as defined by 3GPP. Features that enable accurate comparison between the two technologies e.g., Adaptive Coding and Modulation (ACM) and similar scheduling and Radio Resource Management (RRM) algorithms, have also been implemented.

### B. 5G NR NTN System Level Simulator – ALIX

ALIX is a 5G (TN-)NTN SLS [19], initially developed as part of the ESA ALIX project [20], with the goal of contributing to the successful standardization of the Non-Terrestrial Network (NTN) segment within 3GPP. Like SNS3, the simulator is an extension of ns-3, with its own L2S mapper for different Modulation and Coding Schemes (MCS). The simulator calculates SINR for each received transport block, including received power, noise and co-channel interference and uses the L2S to convert that into a block error rate (BLER). Similarly to SNS3, obtained LLS results have also been utilized for SINR-to-BLER mapping. To model 5G networks, ns-3 has been extended with 5G-LENA [21], which models physical- and Medium Access Control (MAC) layers of NR and implements terrestrial propagation and channel models of 3GPP TR 38.901 [22]. 5G-LENA implements the channel, NR Physical Layer (PHY) and MAC protocol layers, algorithms, and procedures, but Radio Resource Control (RRC), Packet Data Convergence Protocol (PDCP) and Radio Link Control (RLC) layers are reused from the ns-3 Long Term Evolution (LTE) module. 5G-LENA focuses only on terrestrial network deployment scenarios. To support NTN-specific features, 5G-LENA and ns-3 have been extended by adding support for 3GPP TR 38.811 [2] based channel and antenna/beam modelling, along with a global coordinate system, and the system level calibration scenarios presented in TR 38.821 [3]. The ns-3 platform provides the higher protocol layers, i.e., network, transport, and application layers. An overview protocol architecture of the ALIX simulator is presented in Fig. 2.

## III. SIMULATION ASSUMPTIONS

This section describes the specifics of the applied comparison scenario and other related assumptions e.g., link budget, traffic models, control channels, Power Control (PC) parameters, DVB-RCS2 frame configurations and RRM algorithms. The simulation scenario is divided into three specific sub-scenarios: 1) *Set-1 Full Load*, which combines the NTN Set-1 link budget parameterization with a simple Full Buffer traffic model; 2) *Set-2 Full Load with Terminal Diversity*, which combines the NTN Set-2 link budget parameterization with Full Buffer traffic model and various user terminal types, resulting in more diverse SINR; and 3) *Set-1 Limited Load*, which uses Set-1 but operates under a lighter load. Together, these sub-scenarios provide a comprehensive overview of different network conditions for the protocols being compared.

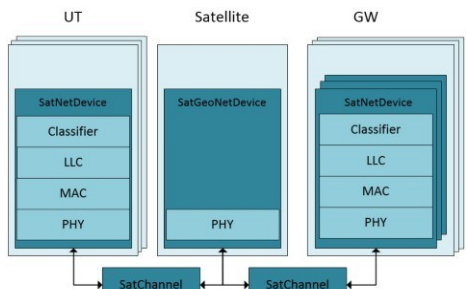

Fig. 1. SNS3 SLS air interface protocol architecture.

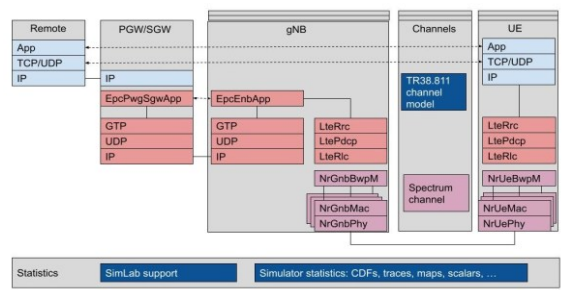

Fig. 2. ALIX SLS protocol architecture.

## A. Simulation scenario

To compare the two technologies, an equivalent scenario setup is provided for both simulators: a single GEO satellite providing service to the same area, with central beam elevation angle of 45 degrees. The GEO satellite is deployed with either Set-1 or Set-2 characteristics, defined in 3GPP TR 38.821 Tables 6.1.1.1-1 and 6.1.1.1-2, depending on the sub-scenario. The central beam is used for collecting statistics and the surrounding beams are used for providing background interference to the simulation. Either 1 or 2 tiers of interfering beams are configured, depending on the sub-scenario design and its computational demand: for instance, the return link and limited-load scenarios require individual user terminals (UTs) to be modelled in detail, while on forward link a fully loaded system is simpler to model with full-band transmissions. The satellite payload architecture is assumed transparent with ideal feeder link, i.e., only user link affects SINR. The scenario is visualized with 2 tiers of interfering beams, and with both Set-1 and Set-2 in Fig. 3 (a) and (b), respectively.

The channel model is configured identically in both simulators, consisting of Line-Of-Sight (LOS) only Free Space Path Loss (FSPL) channel, coupled with 3GPP TR 38.811 antenna gain models and without additional effects, such as shadowing. Additionally, the *Full Load* sub-scenarios have the spatially correlated ITU-R P.1853.2 weather attenuation model [23] enabled to provide variance in the received signal power, at the cost of added computational complexity.

An equal number of 50 stationary Very Small Aperture Terminal (VSAT) UTs are deployed in each beam area with randomized positions to provide reasonable network activity level concerning scheduling and interference. UT and beam positions are consistent between simulators i.e., the same positions are used in specific simulation drops for both simulators, with each drop representing simulations utilizing different Random Number Generator (RNG) realizations. Interfering beams may contain less than 50 UTs, if permissible by the scenario. Statistics are collected only from UTs connected to the central beam.

The frequency division between beams in the scenario follows the FRF-2+2 scheme, defined as Option 3 by 3GPP [3]. Only a single color is simulated, to limit simulation time and memory usage. Additionally, neither adjacent channel interference nor cross-polarization interference is considered. Effectively, interference is only caused by neighboring beams of the same color and intermodulation interference from the satellite and UT power amplifiers.

## B. General Assumptions

The general simulation assumptions are detailed in Table I.

Table I. General simulation assumptions.

| Parameter | Value |
|---|---|
| Simulation duration | 5 seconds per simulated drop (5 drops) |
| Frequency configuration and waveforms | Ka-band: 20 GHz center carrier frequency on DL and 30 GHz on UL. 200 MHz single beam bandwidth per link direction.<br>DVB-S2X: 1 Time Division Multiplexing (TDM) carrier per beam (roll-off = 0.05, carrier spacing 0.02), MODCODs with Quadrature Phase Shift Keying (QPSK), 8-Phase Shift Keying (8PSK), 16-Amplitude and Phase Shift Keying (16APSK) and 32APSK constellations and varying coding rate, Normal FECFRAMEs [5].<br>DVB-RCS2: Static 10 x 20 MHz or 40 x 5 MHz carriers, 12.456 ms superframe duration, waveforms 13-22 [7] (roll-off = 0.20, carrier spacing 0.02).<br>NR: Numerology 3, 132 Physical Resource Blocks (PRB), MCS table 3 [24] |
| ACM | Enabled |
| Automatic Repeat request (ARQ) / Hybrid ARQ (HARQ) | Disabled |
| Control channel and signaling | Ideal (no errors) signaling with delay, no control channel overhead modelled, control components affecting results listed below.<br>NR: Buffer Status Report (BSR) and Downlink Control Information (DCI).<br>DVB: Capacity Request (CR) and Terminal Burst Time Plan (TBTP). |
| Physical layer overheads | DVB-S2X: BaseBand frame (BBframe) header, Physical Layer (PL) header, pilot blocks.<br>DVB-RCS2: Waveform and burst overheads (guard symbols, pilots, pre-amble, post-amble).<br>NR: Demodulation Reference Signal (DMRS), Phase-Tracking Reference Signal (PTRS), Transport Block (TB) overheads i.e., Cyclic Redundancy Check (CRC), code rate |
| Non-linear power amplifier | DVB-S2X/NR Physical Downlink Shared Channel (PDSCH) [9]:<br>Input Back-Off (IBO) = 5.0 dB<br>Output Back-Off (OBO) = 0.8 dB<br>Carrier-to-intermodulation noise ratio (C/Im) = 18.6 dB for DVB-S2X, 18.4 dB for NR.<br>NR Physical Uplink Shared Channel (PUSCH) and DVB-RCS2 [9]:<br>IBO = 0.0 dB<br>OBO and associated C/Im depend on the used MODCOD/MCS and Uplink PC. |
| Scheduler | DVB-S2X/NR PDSCH:<br>PF ($\alpha=0$, $\beta=1$) , using (1)<br>NR PUSCH and DVB-RCS2:<br>Round robin (NR) / Resource-fair (DVB-RCS2)<br>NR: Dynamic multi-user scheduling, UEs multiplexed in the same slot. |
| UL power control | Enabled<br>DVB-RCS2 Energy per symbol to Noise density ratio (Es/N0) and NR PUSCH Signal-to-Noise Ratio (SNR) targets set at 12.5 dB and 15 dB, respectively.<br>(DVB-RCS2: Es/N0 target per transmission [7], NR PUSCH: CLx-ile power control [25]) |
| Channel estimation (NR PUSCH and DVB-RCS2) | Carrier to Noise density ratio (C/N0) and Channel Quality Information (CQI) reporting every configured interval of 100 ms.<br>Minimum SINR value in a moving window with a configured period of 500 ms used as basis for C/N0 and CQI estimation. |
| Traffic models | Full buffer (Radio Link Control (RLC) layer for NR, Logical Link Control (LLC) layer for DVB).<br>3GPP File Transfer Protocol 3 (FTP3) [26], as limited-load traffic model, with below parameters. |
| FTP3 parameters | Mean file inter-arrival time: 100 ms (Poisson distribution)<br>Poisson interval upper bound: 1 s.<br>File size: configurable, 15 kB applied for UL simulations, 60 kB applied for DL, to demonstrate different system loads. |

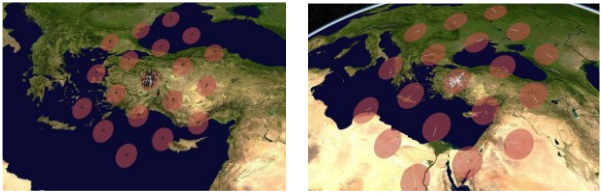

Fig. 3. Scenario visualization, 2 tiers of interfering beams.

(a) 3GPP Set-1 scenario  (b) 3GPP Set-2 scenario

The DL scheduling assumes a Proportional Fair (PF) scheduler, formulated as: user prioritization metric $P_\text{user}$ is defined as a function of achievable throughput in current time slot $T_\text{user}$, historical average throughput $R_\text{user}$ and fairness indices α, β:

$$P_\text{user} = \frac{T_\text{user}^{\alpha}}{R_\text{user}^{\beta}}. \quad (1)$$

The UL is set to use resource-fair, Round-Robin schedulers. While the schedulers are not identical for the two different protocols, which have entirely different radio frame configurations, the schedulers operate with the same principle of targeting at allocating equal amounts of time-frequency resources for each user. The non-linear power amplifier characteristics have been adapted from [9], with the parameters chosen to reach a good compromise between C/Im and transmission power, to maximize overall SINR.

*C. Terminal characteristics*

The characteristics of UTs within the simulation scenarios are outlined in Table II. In the *Set-1 Full Load* and *Set-1 Limited Load* sub-scenarios, only the default VSAT terminal is used. In contrast, the *Set-2 Full Load with Terminal Diversity* sub-scenario distributes approximately equal proportions of three terminal types within each beam, with two terminal types having 17 terminals each and one terminal type having 16, for a total of 50 users per beam. This distribution ensures a balanced setup for a fair comparison across drops. All UTs are assumed to be perfectly pointed towards the satellite, i.e., no antenna gain loss from pointing error.

## IV. SIMULATION RESULTS

The results obtained using the described simulators under the defined assumptions are detailed below. Results are first split into two categories, based on the link direction (DL or UL). Second, the comparison results in those categories are split into the three sub-scenarios: *Set-1 Full Load* (default VSAT terminals, 3GPP Set-1), *Set-2 Full Load with Terminal Diversity* (three different terminal types, 3GPP Set-2), and *Set-1 Limited Load* (default VSAT terminals, 3GPP Set-1, 3GPP FTP3 [26] traffic model). Lastly, results within the sub-scenario are divided according to the respective technologies used: DVB-S2X vs. NR PDSCH on forward link/DL, and DVB-RCS2 vs. NR PUSCH on return link/UL.

The user-level results are presented in Table III, which contains the SINR and user throughput distributions and comparative average throughput gains. Additionally, the user results are presented in Fig. 4., where (a), (b) and (c) contain the user link SINR, user throughput and user file throughput (defined below) distributions for the DL, and (d), (e) and (f) for the UL, respectively. The results show mostly aligned SINR distributions for corresponding scenarios between forward link protocols, and DVB-S2X providing higher user throughput on average than NR PDSCH. In the return link direction, there is more variation in SINR due to more complex architecture, and NR PUSCH provides higher user throughput on average than DVB-RCS2 in the end. Beam-level PHY spectral efficiency and RLC/LLC layer throughputs are detailed in Table IV, reflecting the user-level findings. All presented results assume Additive White Gaussian Noise (AWGN) receiver performance. The ITU-R P.1853.2 weather attenuation model [23] is assumed to be in use in the *Set-1 Full Load* and *Set-2 Full Load with Terminal Diversity* results, slightly yet fairly affecting the experienced signal strength. Further assumptions are detailed in the footnotes of Table III.

Contrary to the throughput with Full Buffer traffic model, in the limited-load scenario the throughput is evaluated as user file throughput, defined as

$$TP_{file} = \frac{S}{t_{Rx\_finish} - t_{Tx\_start}}, \quad (2)$$

where $S$ is the file size in bits, $t_\text{Rx\_finish}$ is the instance at which the file was received, and $t_\text{Tx\_start}$ is the time instance at which the file transmission started. As a user may transmit multiple files during the statistics collection phase and the circumstances may vary between the files, the throughput distribution is sampled from individual files instead of users.

## V. DISCUSSION

Overall, it was found that DVB-S2X demonstrated better or nearly equal performance throughout the system level comparison. This was largely due to the LLS results [12], which provided higher demand for SINR in NR to reach similar spectral efficiency levels to DVB-S2X, indicative of generally slightly more efficient waveform configurations supported by DVB-S2X. In a fully loaded system, the interference was maximal for both technologies. By applying PF scheduling (i.e. bit-fair) for the UTs, the outcome after all physical level overheads and equivalent system-level overheads from scheduling, multi-user beams and air interface protocols was clearly in favor of DVB-S2X.

Table II. User terminal characteristics.

| Characteristics | VSAT low capability | VSAT (default) | VSAT high capability |
|---|---|---|---|
| Frequency band | Ka-band (20 GHz DL, 30 GHz UL) | Ka-band (20 GHz DL, 30 GHz UL) | Ka-band (20 GHz DL, 30 GHz UL) |
| Antenna type and configuration | Directional with 46 cm equivalent aperture diameter | Directional with 60 cm equivalent aperture diameter | Directional with 180 cm equivalent aperture diameter |
| Polarization | circular | circular | circular |
| Rx antenna gain | 36.7 dBi | 39.7 dBi | 53.2 dBi |
| Antenna temperature | 150 K | 150 K | 150 K |
| Noise figure | 1.2 dB | 1.2 dB | 1.2 dB |
| Tx transmit power | 1 W (30 dBm) | 2 W (33 dBm) | 2 W (33 dBm) |
| Tx antenna gain | 40.4 dBi | 43.2 dBi | 50.1 dBi |

Having only one type of UT (i.e. default 3GPP VSAT) maximized the performance gap in favor of DVB-S2X, while with a larger variety of different terminals, and resulting SINR range, the performance gap was slightly smaller. However, under a more limited system load, the practical performance for delivering upper-layer payloads proved almost equal as the system load decreased. User file throughput, used as a metric for performance evaluation of the limited-load scenario, was highly dependent on the realized protocol and propagation delays due to equalized payload sizes. Therefore, similar performance using this metric indicates that both technologies also have similar delay performance on the DL. This result was largely influenced by DVB-S2X having dummy frames being sent when there was no data to transmit, leading to full interference between the satellite beams, and in comparison, the NR system increasing its SINR in contrast to DVB-S2X.

As for the UL comparison, it was observed that NR PUSCH on average provided noticeably better throughput than DVB-RCS2. In a fully loaded system, with 3GPP Set-1 parameterization and single VSAT terminal type, NR demonstrated better user throughput across the scenario, despite lower average SINR, indicating that NR can reach higher spectral efficiencies with lower SINR demand. The performance gap, and lowered SINR demand, is largely due to NR benefitting more from the relaxed target BLER of $10^{-3}$, compared to DVB-RCS2, which has steeper SINR-to-FER/BLER mapping curves and consequently experiences less benefit from the increase in target error rate. Relaxing the target error rate was justified for this scenario, as the conditions were quite stable, and the realized error rates remained close to, or even below, the given target. The higher

Table III. User SINR and throughput results.

| Link direction | Sub-scenario | Technology | SINR 5th %-ile [dB] | SINR 50th %-ile [dB] | SINR 95th %-ile [dB] | SINR average [dB] | 5th %-ile user tput [kbps] | 50th %-ile user tput [kbps] | 95th %-ile user tput [kbps] | User tput average [kbps] | Average tput gain (vs. comp.) |
|---|---|---|---|---|---|---|---|---|---|---|---|
| DL[a] | Set-1 Full Load[c] | DVB-S2X | 7.4 | 7.9 | 8.1 | 7.8 | 8331.3 | 8458.4 | 8830.1 | 8489.2 | +32.8% |
| | | NR PDSCH | 7.4 | 7.9 | 8.1 | 7.8 | 6389.8 | 6389.8 | 6389.8 | 6390.6 | -24.7% |
| | Set-2 Full Load with Terminal Diversity[d] | DVB-S2X | 0.2 | 3.4 | 9.0 | 4.2 | 4798.2 | 5165.0 | 5330.3 | 5155.9 | +21.7% |
| | | NR PDSCH | 0.3 | 3.1 | 9.0 | 3.6 | 3719.1 | 4055.1 | 4865.0 | 4235.6 | -17.8% |
| | Set-1 Limited Load[d] | DVB-S2X | 8.3 | 8.5 | 8.7 | 8.5 | 1745.9 | 1761.3 | 1764.3 | 1759.3 | +4.9% |
| | | NR PDSCH | 8.4 | 9.5 | 10.9 | 9.5 | 1485.3 | 1692.8 | 1760.5 | 1676.8 | -4.7% |
| UL | Set-1 Full Load[b,d] | DVB-RCS2[e] | 8.8 | 10.4 | 12.3 | 10.5 | 5521.6 | 6174.7 | 6339.0 | 5970.3 | -20.0% |
| | | NR PUSCH | 9.1 | 9.6 | 10.4 | 9.6 | 6781.7 | 7538.1 | 8191.5 | 7466.6 | +25.0% |
| | Set-2 Full Load with Terminal Diversity[a,d] | DVB-RCS2[e] | 2.3 | 9.0 | 14.6 | 8.8 | 1876.7 | 5532.8 | 9523.1 | 5851.7 | -14.9% |
| | | NR PUSCH | 5.7 | 9.6 | 11.2 | 9.2 | 956.2 | 4661.6 | 14731.4 | 6872.6 | +17.4% |
| | Set-1 Limited Load[a,d] | DVB-RCS2[f] | 11.1 | 13.3 | 14.8 | 13.2 | 150.3 | 273.0 | 415.8 | 275.8 | -23.3% |
| | | NR PUSCH | 9.0 | 9.9 | 12.3 | 10.2 | 219.9 | 385.3 | 432.7 | 359.5 | +30.3% |

a. BLER target =1E-5.
b. BLER target =1E-3.
c. 2 interfering tiers of beams.
d. 1 interfering tier of beams.
e. Static frame configuration 40 X 5 MHz.
f. Static frame configuration 10 X 20 MHz.

Table IV. Beam spectral efficiency and throughput results.

| Link direction | Sub-scenario | Technology | Average beam PHY spectral efficiency over system bandwidth (400 MHz) [b/s/Hz] | Average beam PHY spectral efficiency (200 MHz) [b/s/Hz] | Average beam throughput (on RLC/LLC) [Mbps] | Average throughput gain (vs. comparison) |
|---|---|---|---|---|---|---|
| DL | Set-1 Full Load | DVB-S2X | 1.062 | 2.124 | 424.5 | +32.9% |
| | | NR PDSCH | 0.799 | 1.598 | 319.5 | -24.7% |
| | Set-2 Full Load with Terminal Diversity | DVB-S2X | 0.640 | 1.280 | 255.8 | +20.8% |
| | | NR PDSCH | 0.530 | 1.060 | 211.8 | -17.2% |
| UL | Set-1 Full Load | DVB-RCS2 | 0.747 | 1.494 | 298.5 | -20.0% |
| | | NR PUSCH | 0.954 | 1.908 | 373.3 | +25.1% |
| | Set-2 Full Load with Terminal Diversity | DVB-RCS2 | 0.732 | 1.464 | 292.6 | -14.8% |
| | | NR PUSCH | 0.872 | 1.744 | 343.6 | +17.4% |

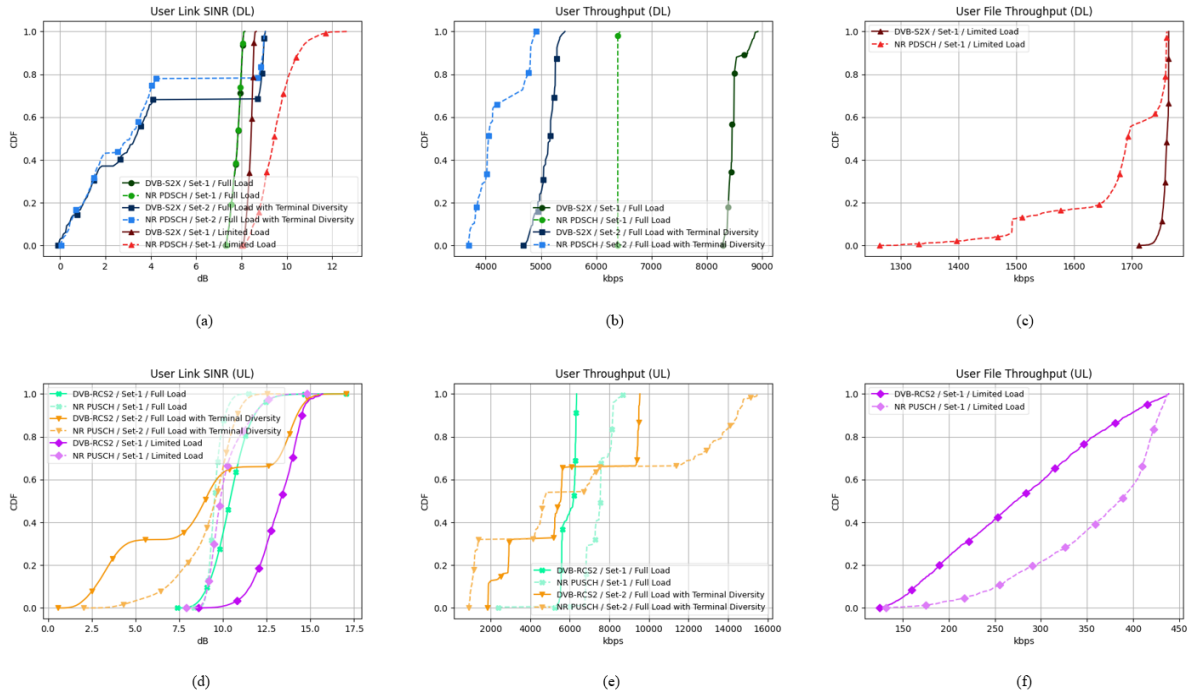

Fig. 4. User SINR and throughput results; cumulative distributions.

SINR of DVB-RCS2 was largely due to the power control methodology, which was based on Es/N0 measurements rather than on SNR, as in NR.

Extending the scenario to utilize 3GPP Set-2 and three types of VSAT terminals with different capabilities demonstrated a diminishing average performance gain for NR PUSCH. This was primarily due to the dynamic nature of resource allocation in NR, which assigned smaller allocations for lower-capability terminals, partly due to the PC limiting the allocation size. Combined with robust MCS selection, this results in reduced throughput for low-capability terminals, while freeing up resources for high-capability terminals, leading to significantly higher peak throughput. On the other hand, DVB-RCS2 provided improved throughput for low-capability terminals largely due to the static frame configuration, i.e., the same bandwidth resources were allocated to every UT.

NR PUSCH was observed to provide the highest gains in the limited-load scenario comparison, where delay was a critical aspect of the evaluated file throughput metric, indicating that NR performed better in minimizing the realized delay on upper-layer payloads within this scenario. While the overall throughput of lower-layer packets reached a very similar performance in lower system loads, DVB-RCS2 struggled to reach the same upper-layer packet delay performance compared to NR, and in comparison, delivered lower performance in a bursty traffic scenario. This was partially due to NR being able to dynamically allocate resources based on need, while DVB-RCS2 used a static frame allocation, resulting in comparatively lower spectral occupancy and higher file transmission durations on average for DVB-RCS2. Due to the lower spectral occupancy, interference in the DVB-RCS2 system was lower in general, enabling also higher SINR; however, DVB-RCS2 could not fully utilize the improved SINR due to already using the most efficient waveform.

Reflecting these results on earlier research is an important step in validating whether the results are representative and result in reasonable conclusions. In [10], the spectral efficiency and forward error correction (FEC) coding performance of 5G NR, DVB-S2X, DVB-RCS2 and the proprietary Mx-DMA Multi Resolution Coding (MRC) waveforms were compared. The study demonstrated that 5G NR could potentially reach similar FEC coding performance, compared to the other waveforms. As for the spectral efficiency, 5G NR and DVB-RCS2 were observed to perform similarly, assuming a resource block (RB) allocation of 5 or more for NR. Comparing NR to either DVB-S2X or Mx-DMA MRC, the spectral efficiency was reported to be around 15% worse for NR, assuming a high bandwidth (20 MHz). Comparing these observations with the supporting link level study [12], from which the link level performance was incorporated into this system level evaluation, the studies seem to arrive at similar conclusions. In [12], it was observed that the spectral efficiency loss and Signal-to-noise Ratio (SNR) degradation from framing, spectral occupancy, implementation losses, and impairments are significantly lower for DVB-S2X compared to NR PDSCH, whereas DVB-RCS2 and NR PUSCH have similar performance. Moreover, the results showed that the demodulation performance of DVB-S2X was noticeably better than NR PDSCH as well, assuming low FER/BLER targets of either 1e-3 or 1e-5. Using the same error rate targets, NR PUSCH was evaluated to have generally slightly better demodulation performance, compared to DVB-RCS2. In both DL and UL, NR performed comparatively better with an increasing target error rate. Previous system level evaluation activities, functioning as a precursor to this study, have also been previously conducted in [11], utilizing the same simulation tools, although with vastly different assumptions and without the results from [12]. Although evidently less refined and more limited in simulation scope, the system level evaluation arrived in similar conclusions as can be observed from the results here i.e. DVB-S2X performs generally better than NR PDSCH within this GEO, Ka-band, evaluation scenario. System level simulations in UL were not conducted in the evaluation, although analysis of literary sources suggested that DVB-RCS2 might perform marginally worse than NR PUSCH.

Despite the comparison already covering multiple scenarios, different system loads, and the effect of multiple terminal types, the comparison could be further extended. For one, the scenario geometry and assumptions could be modified to reach even higher and/or lower ranges of SINR. The weather attenuation model was considered to have a relatively low effect on SINR, as the scenario location was within an area with a low rainfall probability in the methodology for rainfall prediction in [23], thus the probability of experiencing deep fading was low, and the spatial consistency ensured similar conditions across the same geographical areas; however, weather effects in different geographical areas could still contribute to an increased variety of SINR. Moreover, this comparison was focused on the GEO, Ka-band environment, and further comparisons in different settings, such as with a single Non-Geostationary Orbit (NGSO) satellite or NGSO constellations, varying frequency bands, bandwidth limitations, and mobility, could prove worthwhile. This activity also left out the effects of upper-layer protocols, such as Internet Protocol (IP), Transmission Control Protocol (TCP), User Datagram Protocol (UDP) and application layer protocols, which could be evaluated in further activities to provide a more complete view of the performance of the respective air interface technologies. The effects and features of control plane signaling were also mostly left out of the study, with control signals being largely idealized, as for NR the control plane is dependent on network configuration and mobility/cell selection-triggered RRC messages with various capacity demands, and for DVB-S2X/RCS2 the control plane is left for the network implementation (being unspecified). Therefore, considering the effects of the control plane makes it challenging to directly compare the technologies in a fair manner.

Following the recent expansion of the DVB-RCS2 standard into NGSO in [27], along with other potentially performance enhancing revisions, an increasingly relevant direction for future evaluation would be to compare the return links of NR NTN and DVB-RCS2 in NGSO. The updated specification also allows the usage of smaller waveform roll-off factors, namely 0.05 and 0.10, than previously specified, reducing the physical layer overhead, as well as implementing the DVB-S2X waveforms in the return link, which have the potential to perform better than NR NTN in certain scenarios, as was observed within this activity. Reducing the DVB-RCS2 roll-off factor could be an obvious improvement to the comparison within this activity as well, although it was decided to prefer maintaining consistency with the related LLS activity, which utilized a roll-off factor of 0.20, as well as the results reported to ETSI SES-SCN [9].

Evidently, the comparison depends in part on the specifics of the simulation assumptions, and even small changes can lead to significant differences in the results. The simulation approach of this comparison could be further optimized or adjusted, for example, through modifications to power amplifier modelling, power control schemes and parameters, or the RRM algorithms used.

## VI. CONCLUSIONS

This article presents the results of a simulative comparison between DVB-S2X/RCS2 and NR NTN in a GEO satellite scenario utilizing Ka-band frequencies. The comparison utilizes NTN-specific example parameterizations defined by 3GPP to ensure a fair evaluation of the technologies. The underlying effects and causes behind the results, as well as potential directions for further comparisons, are briefly discussed. The SNS3 and ALIX system level simulators behind these results are concisely described, along with the relevant simulation assumptions and parameters.

The results, obtained under given assumptions, indicate that DVB-S2X performs generally better compared to NR PDSCH in the DL direction. However, under lower system loads, the performance of both technologies is almost equal, with their performance evening out as system load decreases. Contrary to observations in the DL direction, in the UL direction NR PUSCH performs generally better than DVB-RCS2 across the evaluated scenarios. This performance gap is especially pronounced with lower system loads, with delay-dependent upper-layer payload throughput being evaluated.

ACKNOWLEDGMENT

This activity was carried out under a programme of, and funded by, the European Space Agency in the scope of ARTES 4.0 project "Multi Access RelatIve performaNce compArison for GSO broadband satellite networks (MARINA)", contract no 4000141353/23/NL/FGL. The authors would like to acknowledge the colleagues at Thales Alenia Space France, Baptiste Chamaillard, Albekaye Traoré and Nathan Borios for link-level analysis and support.